\documentclass[12pt]{article}
\usepackage{cmap}
\usepackage[latin1]{inputenc}
\usepackage[T1]{fontenc}
\usepackage[british]{babel}

\usepackage{lmodern}

\usepackage{amssymb, amsmath, amsthm}
\usepackage[a4paper,top=25mm,bottom=25mm,left=25mm,right=25mm]{geometry}
\usepackage{ragged2e}

\usepackage{authblk} 
\usepackage{pifont}
\usepackage{graphicx}
\usepackage[usenames,dvipsnames,svgnames,table]{xcolor}
\usepackage[figuresright]{rotating}
\usepackage{xtab} 
\usepackage{longtable} 
\usepackage{multirow}
\usepackage{footnote}
\usepackage[stable]{footmisc}
\usepackage{chngpage} 
\usepackage{pdflscape} 
\usepackage[nottoc,notlot,notlof]{tocbibind} 

\usepackage{pgfplots}
\pgfplotsset{every tick label/.append style={font=\footnotesize}}
\pgfplotsset{compat=1.14}
\usepackage{setspace}

\usepackage{array}
\newcolumntype{K}[1]{>{\centering\arraybackslash$}p{#1}<{$}}

\makesavenoteenv{tabular}
\usepackage{tabularx}
\usepackage{booktabs}
\usepackage{threeparttable}
\usepackage[referable]{threeparttablex} 
\newcolumntype{R}{>{\raggedleft\arraybackslash}X}
\newcolumntype{L}{>{\raggedright\arraybackslash}X}
\newcolumntype{C}{>{\centering\arraybackslash}X}
\newcolumntype{A}{>{\columncolor{gray!25}}C}
\newcolumntype{a}{>{\columncolor{gray!25}}c}

\newlength{\tablen}

\usepackage{dcolumn} 
\newcolumntype{.}{D{.}{.}{-1}}

\usepackage{tikz}
\usetikzlibrary{arrows, calc, matrix, patterns, positioning, trees}
\usepackage[semicolon]{natbib}
\usepackage[hyphens]{url}
\usepackage{hyperref} 
\hypersetup{
  colorlinks   = true,    		
  urlcolor     = blue,    		
  linkcolor    = blue,    		
  citecolor    = ForestGreen	
}
\usepackage{microtype}
\usepackage[justification=centering]{caption} 

\usepackage[labelformat=simple]{subcaption}

\DeclareCaptionLabelFormat{parenthesis}{(#2)}
\makeatletter
\renewcommand\p@subfigure{\arabic{figure}.}
\makeatother

\DeclareCaptionLabelFormat{parenthesis}{(#2)}
\makeatletter
\renewcommand\p@subtable{\arabic{table}.}
\makeatother

\usepackage{enumitem}

\setlist[itemize]{leftmargin=2.5\parindent}
\setlist[enumerate]{leftmargin=2.5\parindent}

\def\addlegendimage{\csname pgfplots@addlegendimage\endcsname}

\theoremstyle{plain}

\theoremstyle{definition}

\theoremstyle{remark}

\makeatletter
\let\@fnsymbol\@alph
\makeatother

\def\keywords{\vspace{.5em} 
{\noindent \textit{Keywords}: }}

\def\AMS{\vspace{.5em} 
{\noindent \textbf{\emph{MSC} class}: }}

\def\JEL{\vspace{.5em} 
{\noindent \textbf{\emph{JEL} classification number}: }}

\title{The efficacy of tournament designs}
\author{Bal\'{a}zs R. Sziklai\thanks{~Email: \emph{sziklai.balazs@krtk.hu} \newline Centre for Economic and Regional Studies (KRTK), Budapest, Hungary \newline Corvinus University of Budapest (BCE), Department of Operations Research and Actuarial Sciences, Budapest, Hungary}
$\qquad$
P\'{e}ter Bir\'{o}\thanks{~Email: \emph{biro.peter@krtk.hu} \newline Centre for Economic and Regional Studies, Budapest, Hungary \newline Corvinus University of Budapest (BCE), Department of Operations Research and Actuarial Sciences, Budapest, Hungary}
$\qquad$
\href{https://sites.google.com/view/laszlocsato}{L\'aszl\'o Csat\'o}\thanks{~Corresponding author. Email: \emph{laszlo.csato@sztaki.hu} \newline
Institute for Computer Science and Control (SZTAKI), E\"otv\"os Lor\'and Research Network (ELKH), Laboratory on Engineering and Management Intelligence, Research Group of Operations Research and Decision Systems, Budapest, Hungary \newline
Corvinus University of Budapest (BCE), Department of Operations Research and Actuarial Sciences, Budapest, Hungary}} 
\date{\today}

\begin{document}

\maketitle
\thispagestyle{empty}

\begin{abstract}
\noindent
Tournaments are a widely used mechanism to rank alternatives in a noisy environment. This paper investigates a fundamental issue of economics in tournament design: what is the best usage of limited resources, that is, how should the alternatives be compared pairwise to best approximate their true but latent ranking. We consider various formats including knockout tournaments, multi-stage championships consisting of round-robin groups followed by single elimination, and the Swiss-system. They are evaluated via Monte-Carlo simulations under six different assumptions on winning probabilities. 
Comparing the same pair of alternatives multiple times turns out to be an inefficacious policy.  While seeding can increase the efficacy of the knockout and group-based designs, its influence remains marginal unless one has an unrealistically good estimation on the true ranking of the players. 
The Swiss-system is found to be the most accurate among all these tournament formats, especially in its ability to rank all participants. A possible explanation is that it does not eliminate a player after a single loss, while it takes the history of the comparisons into account.
The results can be especially interesting for emerging esports, where the tournament designs are not yet solidified.


\AMS{62F07, 68U20}

\JEL{C44, C63, Z20}
\end{abstract}

\clearpage

\section{Introduction}

We study the following \emph{ranking problem}. There is a set of alternatives (``players''), characterised by a single attribute (``strength''). The decision-maker does not know the true strengths but can observe the outcome of paired comparisons (``matches'') between any two players. The results of these clashes are noisy in the sense that a stronger player does not always defeat a weaker one, however, the winning probability monotonically increases as a function of the difference in ability.
The aim is to rank \emph{all} players according to their strengths as well as possible. However, we do not consider ties and home advantage.

The schedule of paired comparisons is called \emph{tournament format}. The economics, management science, and sports literature mostly discuss the situation when the players choose the intensity of their effort, and the principal's objective is to provide incentives for achieving its goal(s) \citep{LazearRosen1981, Rosen1986, Taylor1995, Prendergast1999, Szymanski2003, OrrisonSchotterWeigelt2004, BrownMinor2014, BimpikisEhsaniMostagir2019}.
On the contrary, here we consider the level of effort to be fixed: it is assumed that all players always perform at their best. This can be a realistic hypothesis in many high-stake environments like crowdsourcing contests \citep{HouZhang2021}, elections \citep{KlumppPolborn2006}, innovation races \citep{HarrisVickers1987, Yucesan2013, AlesChoKorpeoglu2017}, musical competitions \citep{GinsburghVanOurs2003}, or sports tournaments \citep{Palacios-HuertaVolij2009}.

There are two basic tournament formats.
In the binary, knockout, or single-elimination (henceforth knockout) tournament, the loser of any match is immediately eliminated and cannot be the winner. The selection efficiency of this design has been extensively discussed in economics and statistics, especially concerning the effects of its seeding procedure \citep{Hartigan1968, Israel1981, Hwang1982, HorenRiezman1985, KnuthLossers1987, ChenHwang1988, Edwards1998, Schwenk2000, Glickman2008, VuShoham2011, GrohMoldovanuSelaSunde2012, PrinceColeSmithGeunes2013, Krakel2014, HennessyGlickman2016, Karpov2016, AdlerCaoKarpPekozRoss2017, DagaevSuzdaltsev2018, Karpov2018, ArlegiDimitrov2020, KulhanekPonomarenko2020, Arlegi2021}.
The second prominent format is the round-robin, in which all players face all the others and the players are ranked according to their results \citep{HararyMoser1966, Rubinstein1980}.

\emph{Efficacy} can be defined as the capability of a tournament to reproduce the ranking of the players according to their strength. In the real-world, this ranking is naturally hidden, although there are good proxy measures (e.g.\ Elo scores) based on past performance. Here we assume that the power ranking is known and the matches are decided accordingly.

The exact probability $p_{ij}$ of player $i$ finishing in place $j$ can be derived for a small number of players \citep{David1959, Glenn1960, Searls1963}. However, this approach becomes impossible for tournaments with a large number of matches and a complicated branch structure. For instance, there are $2^{n(n-1)/2}$ possible outcomes in a round-robin tournament with $n$ players.

Therefore, we estimate the probabilities by a Monte Carlo simulation, which is a standard methodology in the literature.
\citet{Appleton1995} aims to determine the chance that the best player wins a particular competition, including random and seeded knockout, round-robin, draw and process, and Swiss-system. 
\citet{McGarrySchutz1997} reveal the ranking efficacy of some traditional tournament structures for eight players under a variety of initial conditions.
\citet{MendoncaRaghavachari2000} compare multiple round-robin tournament ranking methods with respect to their ability to replicate the true rank order of players' strengths.
\citet{Marchand2002} computes the chances of a top-seeded player winning a standard and a random knockout tournament, and gives evidence that the outcome of the two ``antagonistic'' versions may not vary as much as expected.
According to \citet{RyvkinOrtmann2008}, the predictive power---the probability of selecting the best player as the winner---of knockout and round-robin tournaments exhibits non-monotonicity as a function of the number of players for fat-tailed distributions of abilities.
\citet{Ryvkin2010} explores two alternative measures of selection efficiency of these mechanisms, the expected ability of the winner and the expected rank of the winner.

Further studies approach the problem from the perspective of sports.
\citet{ScarfYusofBilbao2009} provide a comprehensive overview of tournament formats used in practice and present an extensive list of metrics, as well as a simulation framework.
\citet{ScarfYusof2011} continue this analysis by examining the effect of seeding policy on outcome uncertainty.
\citet{GoossensBelienSpieksma2012} compare four league formats that have been considered by the Belgian Football Association.
\citet{AnnisWu2006} assess the relative merits of playoff scenarios for NCAA I-A football, which differ in the number, selection, and seeding of playoff teams.
\citet{LasekGagolewski2018} investigate the efficacy of league formats used in the majority of European top-tier association football competitions.
\citet{Csato2021b} analyses four hybrid designs, consisting of knockout and round-robin stages, used in the recent IHF World Men's Handball Championships.

Generally, the simulations reinforce the statistical principle that a larger sample (more matches played) leads to better estimates \citep{LasekGagolewski2018, Csato2021b}.
However, none of these works have addressed explicitly a fundamental issue of economics: what is the best usage of limited resources, i.e.\ which format should be followed if the aim is to approximate the true ranking with a given number of costly matches. Furthermore, except for \citet{Appleton1995}, previous works have not examined the Swiss-system.

It is important to note here that the choice of tournament format is driven by a variety of factors such as fairness \citep{CeaDuranGuajardoSureSiebertZamorano2020, ChaterArrondelGayantLaslier2021, DuranGuajardoSaure2017, GoossensSpieksma2012b, Guyon2015a, Guyon2018a, Guyon2020a, KendallKnustRibeiroUrrutia2010, KendallLenten2017, LalienaLopez2019, VanBulckGoossens2020, Wright2014}, incentive compatibility \citep{Csato2020c, Csato2021a, Csato2022a, DagaevSonin2018, Pauly2014, PrestonSzymanski2003, Vong2017}, maximising attendance \citep{Krumer2020a}, or minimising rest mismatches \citep{AtanCavdaroglu2018}. Nonetheless, accuracy in the ranking of the competitors is clearly among the most important aspects of tournament design since in several sports, broadcasting revenues are distributed on the basis of the teams' final position in the ranking \citep{BergantinosMoreno-Ternero2020a, PetroczyCsato2021}.

\citet{Appleton1995} and \citet{McGarrySchutz1997} have studied tournaments with eight or $16$ players. This paper, similarly to \citet{ScarfYusofBilbao2009}, analyses a tournament with $32$ competitors. The enlargement allows considering a broader range of structures, e.g.\ multi-stage tournament with eight groups. Since the number of competitors in several contests is larger than $16$, this modification strengthens the applicability of our results, too.

Our major contribution resides in showing that the Swiss-system is basically more efficacious than any other tournament format containing the same number of matches, especially with respect to its ability to reproduce the \emph{full} ranking of the players. On the other hand, increasing the number of matches between the same players seems to be wasteful: it is better to integrate two separate knockout contests in an ingenious way than to organise more matches according to the same schedule. Seeding can substantially improve the efficacy of a format, however, it requires an unrealistically good prediction about the true ranking of the players. Real data suggest that actual performance in a tournament can significantly differ from the past performance of the players, leaving seeding a significant, but ultimately marginal element in increasing the accuracy of a format. This is reinforced by a recent statistical study, which reveals that seeding itself does not contribute positively to the success of the teams in the UEFA Champions League and the UEFA Europa League \citep{EngistMerkusSchafmeister2021}. 


\section{Methodology} \label{Sec2}

We test the efficacy of various tournament designs via Monte Carlo simulations.

\subsection{Tournament formats} \label{Sec21}

Deriving a full ranking of the players raises two difficulties. First, certain formats, such as the knockout tournament, do not give a complete order, at least in their traditional configuration. Second, all ties should be resolved.
The first issue is handled by organising extra matches between the players who are already eliminated, thus each competitor plays the same number of matches in each tournament format. For tie-breaking purposes, the traditional Sonneborn-Berger and Buchholz rules are applied.
The following designs are implemented. 

\textbf{Round-robin:} Each player plays one match with all other players. Ties are resolved by the Sonneborn-Berger score. 

\textbf{Double round-robin:} Players participate in two round-robin tournaments. Ranking is derived from the combined scores, ties are broken again by the Sonneborn-Berger rule. 

\textbf{Knockout:} In each round, players are paired to play a match. The loser is eliminated, while the winner proceeds to the next round. The process is repeated until a sole winner remains, which requires $n$ rounds for $2^n$ players. In our simulation, the eliminated players also enter into a knockout tournament. That is, all players eliminated in the first round continue the competition and will be ranked between the 17th and 32nd places. Analogously, players eliminated in the second round compete further for the places from the 9th to the 16th, and so on. Consequently, each player plays five matches.

\textbf{Triple knockout:} A knockout system where elimination/progression is decided on the basis of three matches instead of only one. That is, the players are paired in each round to play three matches and the player with two wins qualifies for the next round---the third game is played even if it is unnecessary when the same player wins both the first and the second games.

\textbf{Draw and process:} It is a double elimination tournament. The players play two parallel knockout championships, seeded such that any players who clash in the first or second rounds of the first tournament (draw) cannot meet in the second tournament (proceed) until the final or semi-final, respectively. The final ranking is obtained by comparing the outcome of the two parts as follows. 

Suppose that the first $k$ positions of the final ranking have already been determined by looking at the first $m$ positions of the knockout results, thus the player(s) ranked at the $(m+1)$th place in the knockout tournaments are considered. If the same player occupies these positions, it will be the $(k+1)$th in the final ranking. If two different players occupy these positions, there are three different cases.
If both of them have already obtained a rank in the final ranking, the investigation continues with the player(s) ranked $(m+2)$th.
If exactly one of them has already obtained a rank in the final ranking, the other one will be the $(k+1)$th.
Otherwise, if none of them have already obtained a rank, they play a tiebreaker match. The winner will occupy the $(k+1)$th position and the loser will be the $(k+2)$th in the final ranking. \\
This format is applied e.g.\ in croquet \citep[Chapter~F1d: Two-Life Events]{Croquet2021}.

\textbf{Multi-stage tournament with 8 groups:} The players are divided into eight groups of four players each, where they play a round-robin tournament. The top two players with the highest scores from each group advance to a knockout tournament to allocate the first 16 places in the upper bracket, while the bottom two from each group play a knockout tournament in the lower bracket. Ties in the round-robin phase are broken by the Sonneborn-Berger rule. 

\textbf{Multi-stage tournament with 4 groups:} Analogous to the previous format, but now there are four groups composed of eight players each. Again, the upper and the bottom half of the players form two knockout tournaments.  

\textbf{Double group:} Similarly to the multi-stage tournament with four groups, the players are divided into four groups of eight players each, where they play a round-robin tournament. The top four players from each group proceed to the second round robin-phase. Here four four-player groups are formed by taking the best player from one group, the second best from another, the third and the fourth best from the third and fourth group, respectively. That process is repeated three more times. In the end, the four newly formed groups each contain a winner, a runner-up, a third- and a fourth-placed player from the first round-robin phase. In parallel, the losers of the first round-robin phase form another four four-player groups (losing branch). In both round-robin phases, ties are resolved by the Sonneborn-Berger rule. The final ranking is derived by a knockout tournament that follows the round-robin phases. Winners and second place runner-ups of the second round-robin compete for the first 8th position. Third- and fourth-place players compete for the $9$--$16$th positions. The $17$--$24$th and $25$--$32$th positions are determined by the matches of the losing branch.  


\textbf{Swiss-system:} It is a non-eliminating tournament format with a fixed number of rounds. The matching algorithm pairs players with (approximately) the same score in every round but two players cannot meet more than once \citep{BiroFleinerPalincza2017, FuhrlichCsehLenzner2021}. In our simulation, the pairing is based on an integer program that aims to match the players with the highest score first. In particular, the program finds a maximum weight matching on a graph where the nodes represent the players and two nodes are connected if the corresponding players have not played against each other. The weights of the edges are the product of the players' incremented score (current score $+1$). The $+1$ increment is needed to avoid zero weights for players who have not yet won any match. The program yields the same result as the blossom algorithm developed by Jack Edmond.

The winner is the player having the most wins at the end. Ties are resolved after the final round by the Buchholz scores, which sums the scores of all opponents.
In contrast to the previous designs, the Swiss-system depends on one parameter, the number of rounds. 

This format is commonly used in bridge, chess \citep[Chapter~C04: FIDE Swiss Rules]{FIDE2020}, croquet \citep[Chapter~F3: Swiss Events]{Croquet2021}, Go, and esports \citep{Hearthstone2020}.


\subsection{Match prediction and simulation} \label{Sec22}

A symmetric probability matrix is used: for each pair of players $A$ and $B$, the fixed winning probability $p_{AB} = 1-p_{BA}$ determines the likelihood that player $A$ wins against player $B$. The ranking is transitive, furthermore, if $p_{AB} > 0.5$ and $p_{BC} > 0.5$, then $p_{AC} \geq \max \{ p_{AB}; p_{BC} \}$.
 
\citet{Appleton1995} and \citet{MendoncaRaghavachari2000} use normally distributed ratings, whereas \citet{McGarrySchutz1997} consider fixed, linearly structured values. Since the number of players is smaller in these articles (eight or 16 in \citet{Appleton1995}, six in \citet{MendoncaRaghavachari2000}, and eight in \citet{McGarrySchutz1997}), their assumptions on winning probabilities cannot be uniquely extended to our case.
\citet{ScarfYusofBilbao2009} use historical data from the UEFA Champions League to predict the outcome of matches, thus their simulation model depends on the sport considered to a large extent.
Therefore, we have decided to examine three theoretical and three real-world scenarios, consequently, all simulations are carried out with six different sets of winning probabilities.  

For each scenario, we conducted 1 million simulation runs. The metrics presented in the following are derived from these simulations. According to the Central Limit Theorem, the sample means converge. To show that the convergence is sufficiently fast, we cut the 1 million simulation runs into 10 equal sized runs and tested the obtained sample means with Student's t-test. At the significance level of 1\%, the sample mean of each metric presented here does not differ from the true value by more than $0.5$\%.  

In the three theoretical scenarios, the probability that player $A$ wins against player $B$ is given by
\[
P_{AB} = 0.5 + \texttt{skill} \times (\texttt{rank(B)}-\texttt{rank(A)})/100,
\]
where $\texttt{skill}$ is a parameter and \texttt{rank(A)} and \texttt{rank(B)} denote the positions of the two players in the real power ranking, respectively. In the first theoretical scenario, even the weakest player has a reasonable chance to win against the strongest player ($\texttt{skill}=1$). In the third, only top players can compete with the best ($\texttt{skill}=10$). The middle models a transition between these two extremities ($\texttt{skill}=5$). 


For instance, if $A$ is the strongest player and $B$ is the second one under $\texttt{skill}=10$, then the former wins with a probability of 60\%. On the other hand, the winning probability of player $A$ is reduced to 51\% if $\texttt{skill} = 1$. If rank difference would indicate a negative probability, it is treated as an impossible event. Similarly, if rank difference leads to a winning probability greater than 1, it is considered to be a sure event.

We also consider real data from three competitions in different sports where the performance of the players can be measured by the Elo scores (note that this methodology is used ``officially'' only in the case of chess):
\begin{itemize}
\item
Chess: the players of the \href{https://en.wikipedia.org/wiki/Chess_World_Cup_2017}{World Chess Cup 2017};
\item
Soccer: the clubs participating in the \href{https://en.wikipedia.org/wiki/2017\%E2\%80\%9318_UEFA_Champions_League}{2017/18 UEFA Champions League};
\item
Tennis: the contestants of the \href{https://en.wikipedia.org/wiki/2017_Monte-Carlo_Rolex_Masters_\%E2\%80\%93_Singles}{2017 Monte-Carlo Rolex Masters -- Singles} tournament.
\end{itemize}
Elo data are obtained from \url{https://en.wikipedia.org/wiki/Chess_World_Cup_2017} for chess, \url{http://clubelo.com/} for soccer (retrieved on 12th September 2017), and \url{http://tennisabstract.com/} for tennis (retrieved 9th April 2017), respectively.
This is a convenient approach for our model because the Elo rating system is explicitly designed to reflect the winning probabilities of the players against each other. Since draws are not allowed, the winning probability of player $A$ against player $B$ is given by the following formula:
\[
P_{AB}=\frac{1}{1+10^{(R_B-R_A)/400}},
\]
where $R_A$ and $R_B$ are the Elo rating of players $A$ and $B$, respectively. 

Heatmaps of the winning probabilities in the six scenarios are provided in the Appendix, see Table~\ref{Table_A1}.

\subsection{Tournament metrics} \label{Sec23}

Efficacy is quantified by looking at the differences between the real and observed rankings.
For this purpose, two types of indicators are used, the average rank of the top players and the number of (weighted) inversions.
The former has been applied by \citet{ScarfYusofBilbao2009}. It compares the top $k$ players in the observed ranking to the $k$ strongest players by dividing the sum of the ranks of the players finishing in the top $k$ positions with the theoretical minimum of $1 + 2 + \cdots + k = k(k+1)/2$.

However, if $k=3$, this metric does not differentiate between the rankings $3 \succ 2 \succ 1$ (when the third best player wins the tournament) and $1 \succ 2 \succ 3$ (when the strongest player wins the tournament). Consequently, the number of inversions is also computed: the number of times when a weaker player is ranked above a stronger one.
In the previous example, the number of inversions is equal to $3$ for the ranking $3 \succ 2 \succ 1$, but it is $0$ for the ranking $1 \succ 2 \succ 3$.

Furthermore, people tend to care more about the winners, and the top places of the rankings attract more attention, hence differences from the real power ranking are more noticeable here \citep{Can2014}. Therefore, we also propose a novel weighted inversion metric that sums the reciprocals of the logarithms of disconcordant positions. 

\begin{table}[t]
\centering
\caption{Examples for the calculation of the weighted inversion metric}
\label{Table1}
\begin{threeparttable}
\begin{tabularx}{0.8\textwidth}{cCCC} \toprule
Reference ranking & Ranking A & Ranking B & Ranking C \\ \bottomrule
1                                                                                & \cellcolor{gray!20}2     & 1                             & \cellcolor{gray!20}3     \\
2                                                                                & 1                             & 2                             & \cellcolor{gray!20}5                             \\
3                                                                                & 3                             & 3                             &  1    \\
4                                                                                & 4                             & \cellcolor{gray!20}5     & 4     \\
5                                                                                & 5                             & 4                             & 2\\ \bottomrule                        
\end{tabularx}
\begin{tablenotes} \footnotesize
\item
 Grey cells highlight players that precede their reference rank.
\end{tablenotes}
\end{threeparttable}
\end{table}

Table~\ref{Table1} presents three examples.
A player with reference rank $j$ that is ranked at position $i<j$ adds $\sum_{k=i+1}^j 1 / \ln (k)$ to the value of the metric. In ranking $A$, the only player who is ranked higher than its reference rank is player 2. Since the inversion takes place in the first position, the reciprocals of the logarithms are summed up starting from $i+1=2$ to $j=2$, which is just a one-term sum. On the other hand, two players precede their reference rank in ranking $C$. Player 3 adds $1 / \ln{2} + 1 / \ln{3}$ weight, while Player 5 adds $1 / \ln{3} + 1 / \ln{4} + 1 / \ln{5}$.
Thus the weighted inversions between the three rankings are as follows:

\begin{eqnarray*}
w(A) & = & \frac{1}{\ln{2}}; \\
w(B) & = & \frac{1}{\ln{5}}; \\
w(C) & = & \frac{1}{\ln{2}} + \frac{1}{\ln{3}} + \frac{1}{\ln{3}} + \frac{1}{\ln{4}} + \frac{1}{\ln{5}}. 
\end{eqnarray*}

Note that both rankings $A$ and $B$ contain only one inversion compared to the reference ranking, however, in ranking $A$ this happens to be in the top position, hence it weighs more. This is manifested in $w(A) = 1 / \ln{2} > 1 /\ln{5} = w(B)$.

These measures characterise the efficacy of tournament formats sufficiently well.
A lower value of them is always preferred to a higher one.


\subsection{Seeding} \label{ses:seeding}

To disentangle the effect of seeding from tournament structure, random seeding is considered as our baseline. In each simulation run, a new random order of the players is generated. 

The impact of seeding is investigated in two ways. To uncover the maximum possible effect of seeding, the true ranking of the players is assumed to be known and they are seeded according to this ranking. Therefore, the seeding is the same in each simulation run.

In particular, the standard seeding is used for the knockout, triple knockout, and draw and process formats. This seeding method is often used in practice and has been extensively studied in the literature (see, for example, \citet{Hwang1982} or \citet{Schwenk2000}). Analogously, the traditional method is applied for the group stage in any tournament design. If there are $k$ groups and $gk$ teams, the $k$ highest-ranked players are placed in pot 1, the next $k$ in pot 2, and so on, hence pot $g$ contains the $k$ lowest-ranked players. After that, each group gets one player from each pot randomly.

Finally, we estimate the expected impact of seeding. Before the players are seeded, their power ranking is perturbed because the organisers do not know the true ranking, they only have a guess based on the previous matches of the players.

In chess, tournament performances are routinely measured since rating performance can be used for tie-breaking. The results of six chess tournaments were taken into account:
\begin{itemize}
\item
European Individual Chess Championship 2017;
\item
European Individual Chess Championship 2018;
\item
European Individual Chess Championship 2019;
\item
Isle of Man 2017 Open -- Masters;
\item
Gibraltar International Chess Festival 2019 -- Masters;
\item
Grand Swiss 2019.
\end{itemize}

Data was gathered from the website \url{chess-results.com}. All six events were Swiss-system tournaments with 9, 10, or 11 rounds. For each tournament, the performances of the top 150 players are considered, leading to 900 data points. The majority of the players were grand masters (GM). For each tournament, the initial rating and the rating performances of the players were used: their difference indicates how the actual and past performances differ. Tournament performance was calculated according to the FIDE regulations. 

\begin{figure}[t!]
\centering

\begin{tikzpicture}
\begin{axis}[
name = axis1,
width = 0.96\textwidth, 
height = 0.6\textwidth,
xlabel = {Player ID},
x label style = {font = \small},
ylabel = {Difference in rating points},
y label style = {font = \small},
xmin = 0,
xmax = 900,
xmajorgrids = true,
log ticks with fixed point,
x tick label style={/pgf/number format/1000 sep=\,},
ymajorgrids = true,
]
\draw[very thick](axis cs:\pgfkeysvalueof{/pgfplots/xmin},0)  -- (axis cs:\pgfkeysvalueof{/pgfplots/xmax},0);
\addplot [blue, only marks, mark = x, mark size = 1pt, ultra thick] coordinates{
(1,-259)
(2,-244)
(3,-232)
(4,-220)
(5,-218)
(6,-206)
(7,-206)
(8,-204)
(9,-200)
(10,-200)
(11,-198)
(12,-192)
(13,-190)
(14,-189)
(15,-188)
(16,-187)
(17,-184)
(18,-182)
(19,-179)
(20,-178)
(21,-177)
(22,-176)
(23,-174)
(24,-174)
(25,-173)
(26,-172)
(27,-169)
(28,-168)
(29,-168)
(30,-167)
(31,-167)
(32,-165)
(33,-163)
(34,-160)
(35,-159)
(36,-158)
(37,-158)
(38,-158)
(39,-156)
(40,-156)
(41,-153)
(42,-153)
(43,-149)
(44,-149)
(45,-149)
(46,-149)
(47,-148)
(48,-147)
(49,-146)
(50,-145)
(51,-145)
(52,-144)
(53,-144)
(54,-143)
(55,-142)
(56,-140)
(57,-139)
(58,-138)
(59,-138)
(60,-137)
(61,-136)
(62,-131)
(63,-129)
(64,-129)
(65,-128)
(66,-128)
(67,-128)
(68,-128)
(69,-127)
(70,-125)
(71,-125)
(72,-125)
(73,-124)
(74,-124)
(75,-121)
(76,-121)
(77,-120)
(78,-120)
(79,-120)
(80,-119)
(81,-118)
(82,-118)
(83,-116)
(84,-116)
(85,-116)
(86,-116)
(87,-114)
(88,-113)
(89,-112)
(90,-110)
(91,-110)
(92,-109)
(93,-109)
(94,-109)
(95,-108)
(96,-107)
(97,-107)
(98,-107)
(99,-106)
(100,-106)
(101,-105)
(102,-105)
(103,-104)
(104,-104)
(105,-103)
(106,-103)
(107,-103)
(108,-102)
(109,-102)
(110,-102)
(111,-102)
(112,-101)
(113,-101)
(114,-101)
(115,-100)
(116,-100)
(117,-99)
(118,-98)
(119,-98)
(120,-97)
(121,-96)
(122,-95)
(123,-95)
(124,-94)
(125,-94)
(126,-94)
(127,-93)
(128,-93)
(129,-93)
(130,-92)
(131,-92)
(132,-92)
(133,-91)
(134,-90)
(135,-89)
(136,-89)
(137,-89)
(138,-89)
(139,-87)
(140,-87)
(141,-87)
(142,-87)
(143,-86)
(144,-84)
(145,-83)
(146,-83)
(147,-83)
(148,-82)
(149,-82)
(150,-82)
(151,-82)
(152,-82)
(153,-81)
(154,-81)
(155,-81)
(156,-81)
(157,-81)
(158,-80)
(159,-80)
(160,-80)
(161,-80)
(162,-79)
(163,-79)
(164,-79)
(165,-79)
(166,-79)
(167,-78)
(168,-77)
(169,-77)
(170,-76)
(171,-76)
(172,-75)
(173,-75)
(174,-75)
(175,-75)
(176,-75)
(177,-75)
(178,-74)
(179,-74)
(180,-74)
(181,-73)
(182,-73)
(183,-73)
(184,-72)
(185,-72)
(186,-71)
(187,-71)
(188,-71)
(189,-70)
(190,-70)
(191,-70)
(192,-70)
(193,-69)
(194,-69)
(195,-69)
(196,-69)
(197,-69)
(198,-68)
(199,-68)
(200,-68)
(201,-67)
(202,-67)
(203,-67)
(204,-66)
(205,-65)
(206,-65)
(207,-64)
(208,-64)
(209,-64)
(210,-64)
(211,-63)
(212,-62)
(213,-62)
(214,-61)
(215,-61)
(216,-61)
(217,-61)
(218,-61)
(219,-61)
(220,-60)
(221,-60)
(222,-60)
(223,-60)
(224,-59)
(225,-59)
(226,-59)
(227,-59)
(228,-59)
(229,-59)
(230,-59)
(231,-59)
(232,-59)
(233,-59)
(234,-58)
(235,-58)
(236,-58)
(237,-58)
(238,-57)
(239,-57)
(240,-57)
(241,-56)
(242,-56)
(243,-56)
(244,-56)
(245,-55)
(246,-55)
(247,-55)
(248,-55)
(249,-55)
(250,-55)
(251,-54)
(252,-54)
(253,-54)
(254,-54)
(255,-54)
(256,-54)
(257,-53)
(258,-53)
(259,-52)
(260,-52)
(261,-52)
(262,-52)
(263,-52)
(264,-51)
(265,-51)
(266,-50)
(267,-50)
(268,-50)
(269,-49)
(270,-49)
(271,-49)
(272,-48)
(273,-48)
(274,-48)
(275,-47)
(276,-47)
(277,-47)
(278,-47)
(279,-46)
(280,-46)
(281,-46)
(282,-46)
(283,-46)
(284,-46)
(285,-45)
(286,-45)
(287,-44)
(288,-44)
(289,-43)
(290,-43)
(291,-43)
(292,-43)
(293,-43)
(294,-43)
(295,-42)
(296,-42)
(297,-42)
(298,-40)
(299,-40)
(300,-40)
(301,-40)
(302,-40)
(303,-40)
(304,-39)
(305,-39)
(306,-39)
(307,-39)
(308,-39)
(309,-38)
(310,-38)
(311,-38)
(312,-38)
(313,-38)
(314,-38)
(315,-37)
(316,-37)
(317,-37)
(318,-37)
(319,-37)
(320,-37)
(321,-37)
(322,-36)
(323,-36)
(324,-36)
(325,-35)
(326,-35)
(327,-35)
(328,-35)
(329,-35)
(330,-34)
(331,-34)
(332,-33)
(333,-33)
(334,-33)
(335,-33)
(336,-33)
(337,-32)
(338,-32)
(339,-32)
(340,-32)
(341,-32)
(342,-32)
(343,-31)
(344,-31)
(345,-30)
(346,-30)
(347,-30)
(348,-30)
(349,-29)
(350,-29)
(351,-29)
(352,-29)
(353,-29)
(354,-29)
(355,-28)
(356,-28)
(357,-27)
(358,-27)
(359,-27)
(360,-27)
(361,-27)
(362,-27)
(363,-27)
(364,-26)
(365,-26)
(366,-26)
(367,-25)
(368,-25)
(369,-25)
(370,-25)
(371,-25)
(372,-25)
(373,-25)
(374,-24)
(375,-24)
(376,-24)
(377,-24)
(378,-24)
(379,-24)
(380,-23)
(381,-23)
(382,-23)
(383,-23)
(384,-23)
(385,-22)
(386,-22)
(387,-22)
(388,-21)
(389,-21)
(390,-21)
(391,-20)
(392,-20)
(393,-20)
(394,-20)
(395,-20)
(396,-20)
(397,-19)
(398,-19)
(399,-19)
(400,-18)
(401,-18)
(402,-18)
(403,-17)
(404,-17)
(405,-16)
(406,-16)
(407,-16)
(408,-16)
(409,-16)
(410,-15)
(411,-15)
(412,-15)
(413,-14)
(414,-14)
(415,-14)
(416,-14)
(417,-14)
(418,-13)
(419,-13)
(420,-13)
(421,-12)
(422,-12)
(423,-12)
(424,-12)
(425,-12)
(426,-11)
(427,-11)
(428,-11)
(429,-11)
(430,-11)
(431,-10)
(432,-10)
(433,-10)
(434,-10)
(435,-9)
(436,-9)
(437,-9)
(438,-9)
(439,-9)
(440,-7)
(441,-7)
(442,-7)
(443,-7)
(444,-7)
(445,-7)
(446,-7)
(447,-6)
(448,-5)
(449,-5)
(450,-4)
(451,-4)
(452,-4)
(453,-4)
(454,-4)
(455,-4)
(456,-4)
(457,-4)
(458,-4)
(459,-3)
(460,-2)
(461,-2)
(462,-2)
(463,-2)
(464,-1)
(465,-1)
(466,-1)
(467,0)
(468,0)
(469,0)
(470,1)
(471,1)
(472,1)
(473,1)
(474,1)
(475,2)
(476,2)
(477,2)
(478,3)
(479,3)
(480,3)
(481,3)
(482,3)
(483,3)
(484,3)
(485,3)
(486,4)
(487,4)
(488,4)
(489,5)
(490,5)
(491,5)
(492,6)
(493,6)
(494,6)
(495,6)
(496,7)
(497,7)
(498,7)
(499,7)
(500,8)
(501,8)
(502,8)
(503,9)
(504,9)
(505,10)
(506,10)
(507,10)
(508,11)
(509,11)
(510,12)
(511,12)
(512,12)
(513,12)
(514,12)
(515,13)
(516,13)
(517,13)
(518,14)
(519,14)
(520,14)
(521,14)
(522,14)
(523,14)
(524,14)
(525,15)
(526,15)
(527,15)
(528,15)
(529,15)
(530,15)
(531,16)
(532,16)
(533,16)
(534,17)
(535,17)
(536,17)
(537,17)
(538,18)
(539,18)
(540,18)
(541,19)
(542,19)
(543,19)
(544,19)
(545,20)
(546,20)
(547,20)
(548,21)
(549,21)
(550,22)
(551,22)
(552,22)
(553,22)
(554,22)
(555,23)
(556,23)
(557,23)
(558,23)
(559,23)
(560,23)
(561,23)
(562,23)
(563,23)
(564,24)
(565,24)
(566,25)
(567,25)
(568,25)
(569,26)
(570,27)
(571,27)
(572,27)
(573,27)
(574,27)
(575,28)
(576,28)
(577,28)
(578,29)
(579,29)
(580,29)
(581,29)
(582,29)
(583,29)
(584,29)
(585,29)
(586,29)
(587,29)
(588,30)
(589,30)
(590,30)
(591,30)
(592,31)
(593,31)
(594,31)
(595,31)
(596,31)
(597,31)
(598,31)
(599,31)
(600,32)
(601,32)
(602,32)
(603,33)
(604,33)
(605,33)
(606,34)
(607,34)
(608,34)
(609,35)
(610,35)
(611,35)
(612,35)
(613,35)
(614,35)
(615,35)
(616,35)
(617,36)
(618,36)
(619,36)
(620,36)
(621,37)
(622,37)
(623,37)
(624,37)
(625,37)
(626,38)
(627,38)
(628,38)
(629,39)
(630,39)
(631,39)
(632,39)
(633,40)
(634,40)
(635,40)
(636,41)
(637,41)
(638,41)
(639,41)
(640,41)
(641,42)
(642,42)
(643,42)
(644,42)
(645,42)
(646,42)
(647,43)
(648,43)
(649,43)
(650,43)
(651,43)
(652,43)
(653,43)
(654,43)
(655,44)
(656,45)
(657,45)
(658,46)
(659,46)
(660,47)
(661,47)
(662,47)
(663,47)
(664,47)
(665,49)
(666,49)
(667,49)
(668,49)
(669,50)
(670,50)
(671,50)
(672,51)
(673,51)
(674,52)
(675,52)
(676,53)
(677,53)
(678,53)
(679,53)
(680,53)
(681,54)
(682,54)
(683,54)
(684,55)
(685,55)
(686,55)
(687,56)
(688,56)
(689,56)
(690,57)
(691,57)
(692,57)
(693,58)
(694,58)
(695,58)
(696,58)
(697,58)
(698,58)
(699,58)
(700,59)
(701,59)
(702,60)
(703,60)
(704,60)
(705,60)
(706,60)
(707,60)
(708,61)
(709,61)
(710,62)
(711,62)
(712,62)
(713,62)
(714,62)
(715,62)
(716,63)
(717,64)
(718,64)
(719,64)
(720,65)
(721,65)
(722,65)
(723,65)
(724,66)
(725,66)
(726,67)
(727,67)
(728,67)
(729,67)
(730,67)
(731,67)
(732,67)
(733,68)
(734,68)
(735,69)
(736,69)
(737,69)
(738,69)
(739,69)
(740,69)
(741,70)
(742,70)
(743,71)
(744,71)
(745,72)
(746,72)
(747,73)
(748,73)
(749,73)
(750,75)
(751,76)
(752,77)
(753,77)
(754,77)
(755,77)
(756,77)
(757,78)
(758,78)
(759,78)
(760,78)
(761,78)
(762,79)
(763,80)
(764,81)
(765,81)
(766,81)
(767,82)
(768,82)
(769,82)
(770,83)
(771,83)
(772,83)
(773,84)
(774,84)
(775,84)
(776,84)
(777,84)
(778,84)
(779,84)
(780,85)
(781,86)
(782,87)
(783,87)
(784,87)
(785,88)
(786,88)
(787,88)
(788,89)
(789,89)
(790,89)
(791,90)
(792,90)
(793,91)
(794,91)
(795,91)
(796,92)
(797,92)
(798,92)
(799,93)
(800,93)
(801,93)
(802,94)
(803,94)
(804,94)
(805,96)
(806,96)
(807,96)
(808,97)
(809,97)
(810,97)
(811,98)
(812,98)
(813,98)
(814,99)
(815,100)
(816,100)
(817,100)
(818,103)
(819,103)
(820,103)
(821,105)
(822,107)
(823,108)
(824,108)
(825,109)
(826,110)
(827,114)
(828,114)
(829,114)
(830,117)
(831,117)
(832,118)
(833,120)
(834,121)
(835,121)
(836,121)
(837,122)
(838,122)
(839,123)
(840,123)
(841,124)
(842,125)
(843,125)
(844,125)
(845,125)
(846,126)
(847,126)
(848,126)
(849,127)
(850,127)
(851,127)
(852,127)
(853,127)
(854,128)
(855,128)
(856,132)
(857,133)
(858,136)
(859,138)
(860,138)
(861,139)
(862,140)
(863,140)
(864,141)
(865,142)
(866,155)
(867,155)
(868,155)
(869,155)
(870,156)
(871,156)
(872,157)
(873,160)
(874,160)
(875,160)
(876,163)
(877,163)
(878,164)
(879,168)
(880,171)
(881,175)
(882,177)
(883,180)
(884,180)
(885,181)
(886,184)
(887,185)
(888,186)
(889,187)
(890,187)
(891,188)
(892,191)
(893,193)
(894,199)
(895,200)
(896,201)
(897,209)
(898,215)
(899,263)
(900,326)
};
\end{axis}
\end{tikzpicture}

\caption{Difference between the initial rating of the players and their performance}
\label{Fig1}

\end{figure}
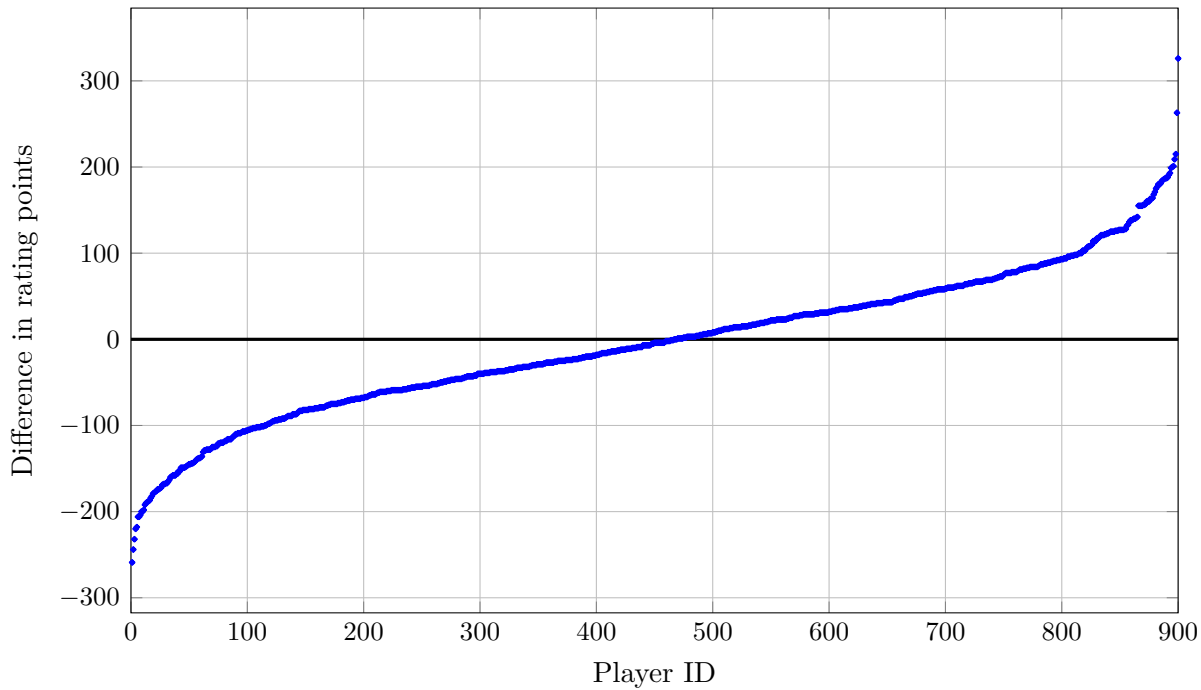



Figure~\ref{Fig1} presents the result. As expected, approximately half of the players underperformed (overperformed) in the tournaments. Furthermore, the difference between the initial and the realised Elo rating is below (above) 100 points for about one-ninth of the players.
To calculate the expected impact of seeding, we start from the real Elo data in the chess scenario. In each simulation run, performance differences are drawn randomly from the empirical distribution given in Figure~\ref{Fig1}. The perturbed ratings represent the real power ranking, while the preliminary ranking is the initial rating used for seeding.


Seeding is not considered for the round-robin and double round-robin designs as each player plays the same number of matches against all the others. Analogously, seeding is not studied for the Swiss-system, where the pairing in each round is determined by the results of the previous rounds, except for the first.

\section{Results} \label{Sec3}

Our discussion begins with the analysis of unseeded tournament formats, followed by examining the effect of seeding and comparing the findings to previous results.

\subsection{Random seeding}
\begin{table}[t!]
  \centering
  \caption{The number of matches in the tournament formats with $32$ players}
  \label{Table2}
    \rowcolors{3}{}{gray!20}
    \begin{tabularx}{0.8\textwidth}{LC} \toprule 
    Tournament format & Number of matches \\ \bottomrule 
    Round-robin & 496 \\
    Double round-robin & 992 \\
    Knockout & 80 \\
    Triple knockout & 240 \\
    Draw and process & 160 \\
    Multi-stage with 8 groups & 112 \\
    Multi-stage with 4 groups & 176 \\
    Double group & 208 \\
    Swiss-system & 16 $\times$ number of rounds \\ \toprule
    \end{tabularx}
\end{table}

According to Table~\ref{Table2}, the number of matches substantially differs for the tournament designs considered.
However, the strength of the players can be estimated only by playing matches, thus the efficacy of any ranking mechanism highly depends on the number of matches played. This is shown in Figures~\ref{Fig2}--\ref{Fig5} by four measures, the number of inversions, the number of weighted inversions, as well as the average rank of the winner (Top 1) and the first eight players (Top 8). For the Swiss-system, the number of rounds is varied between 5 and 14 to contain at least as many matches as the knockout tournament.

\input{Figure2_all_formats_inversions}

\input{Figure3_all_formats_weighted_inversions}

\input{Figure4_all_formats_Top1}

\input{Figure5_all_formats_Top8}

The analysis is centred around two issues:
(1) What is the best way to increase the number of matches if efficacy should be improved? 
(2) How should the competition be designed for a given number of matches?

Concerning the first question, we compare the knockout and round-robin structures, together with the draw and process that consists of essentially two knockout contests. As expected, the two round-robin formats outperform the others.
The gain from playing two round-robin championships instead of one is higher when the competition is less balanced: the number of inversions is reduced by approximately 15\% for our chess data, 25\% for soccer and tennis, as well as for $\texttt{skill}=1$, but 40\% if $\texttt{skill}=10$.

Using a triple knockout system rather than a simple one is clearly uneconomical, the maximal gain in the number of inversions is less than 8\% under all probability models. Integrating two knockout competitions according to the draw and process system is robustly more efficacious than playing three matches for elimination. Interestingly, the relative advantage of draw and process over knockout does not change for its ability to select the winner, which is somewhat counter-intuitive since the latter mechanism is explicitly designed for this purpose.

To conclude, increasing the number of matches in the same design is not a parsimonious way to improve efficacy. Nonetheless, the possibility of ties and home advantage are disregarded in the analysis. Since these factors might be significant (for example, in professional tennis, see \citet{Koning2011}), their incorporation offers an interesting line of future research. 

In order to answer the second question, all formats except for the round-robins can be compared to a Swiss-system tournament with the same number of matches, see Table~\ref{Table2}. In each round of a Swiss-type tournament with 32 players, there are 16 matches. Hence, a 5-round Swiss-tournament has 80 matches just like the simple knockout, whereas a 7-round Swiss competition contains 112 matches, similar to the multi-stage tournament with eight groups. 

The corresponding Swiss-system is robustly preferred to any group-based tournament (multi-stage with 4 or 8 groups, double group). In the case of real data, the advantage of the Swiss-system is found to be higher if the competition is less balanced, that is, in tennis. The gain from the Swiss-system compared to a more simple design is the lowest in chess, which can be surprising because the Swiss-system is applied in this particular sport, showing the strength of traditions.
The tournament measures of the Swiss-system converge to the corresponding measures of round-robin since the latter is equivalent to a Swiss-system where the number of rounds is the number of players minus one.

Note that the metrics of the Swiss-system are non-monotonic as a function of all matches played if $\texttt{skill}=10$.
The ability of the round-robin tournament to select the winner is unexpectedly weak in the view of how the Swiss-system performs under $\texttt{skill}=5$ and $\texttt{skill}=10$.
Finally, the draw and process design is competitive against the Swiss-system if the aim is to determine a sole winner, especially for $\texttt{skill}=10$ when the winning probability of the stronger player often equals one (under this assumption, the best player always wins against the players ranked 6--32). The same observation holds for the knockout format. A possible explanation is that the best player can suffer an unexpected loss, thus playing more matches is favourable for reproducing the full ranking but not necessarily for selecting the winner.

\begin{figure}[t!]
\centering

\begin{tikzpicture}
\begin{axis}[
name = axis1,
width = 1\textwidth, 
height = 0.6\textwidth,
title = {Number of inversions, $\texttt{skill}=5$},
title style = {align=center, font=\small},
xmajorgrids = true,
ymin = 0,
log ticks with fixed point,
x tick label style={/pgf/number format/1000 sep=\,},
ymajorgrids = true,
legend entries = {Swiss-system with five rounds$\qquad$,Knockout},
legend style = {at={(0.5,-0.1)},anchor = north,legend columns = 3,font = \small}
]
\addplot [red, only marks, mark = o, very thick] coordinates{
(29,1)
(30,2)
(31,5)
(32,8)
(33,8)
(34,9)
(35,11)
(36,10)
(37,23)
(38,26)
(39,63)
(40,56)
(41,73)
(42,100)
(43,132)
(44,179)
(45,237)
(46,297)
(47,362)
(48,444)
(49,535)
(50,663)
(51,797)
(52,951)
(53,1082)
(54,1228)
(55,1446)
(56,1654)
(57,1866)
(58,2078)
(59,2245)
(60,2447)
(61,2666)
(62,2773)
(63,3021)
(64,3115)
(65,3230)
(66,3290)
(67,3440)
(68,3511)
(69,3521)
(70,3476)
(71,3621)
(72,3349)
(73,3386)
(74,3286)
(75,3157)
(76,3018)
(77,2788)
(78,2610)
(79,2459)
(80,2316)
(81,2081)
(82,1923)
(83,1845)
(84,1612)
(85,1507)
(86,1294)
(87,1190)
(88,1137)
(89,892)
(90,824)
(91,693)
(92,602)
(93,488)
(94,493)
(95,390)
(96,330)
(97,253)
(98,228)
(99,201)
(100,164)
(101,146)
(102,99)
(103,101)
(104,72)
(105,61)
(106,50)
(107,57)
(108,42)
(109,30)
(110,24)
(111,20)
(112,16)
(113,15)
(114,8)
(115,8)
(116,5)
(117,4)
(118,6)
(119,4)
(120,3)
(121,2)
(122,2)
(123,1)
(124,3)
(127,1)
(128,1)
(135,1)
};
\addplot [blue, only marks, mark = x, mark size = 4pt, ultra thick] coordinates{
(34,1)
(36,1)
(40,3)
(41,1)
(42,2)
(43,2)
(44,3)
(45,10)
(46,5)
(47,7)
(48,8)
(49,16)
(50,20)
(51,29)
(52,33)
(53,38)
(54,52)
(55,66)
(56,74)
(57,102)
(58,114)
(59,128)
(60,186)
(61,208)
(62,238)
(63,291)
(64,283)
(65,357)
(66,379)
(67,456)
(68,502)
(69,555)
(70,632)
(71,716)
(72,863)
(73,904)
(74,981)
(75,1097)
(76,1200)
(77,1324)
(78,1326)
(79,1484)
(80,1564)
(81,1732)
(82,1845)
(83,1872)
(84,1825)
(85,2019)
(86,2064)
(87,2186)
(88,2256)
(89,2178)
(90,2347)
(91,2425)
(92,2309)
(93,2377)
(94,2325)
(95,2406)
(96,2383)
(97,2345)
(98,2399)
(99,2368)
(100,2251)
(101,2256)
(102,2300)
(103,2182)
(104,2136)
(105,2001)
(106,1936)
(107,1905)
(108,1805)
(109,1733)
(110,1688)
(111,1565)
(112,1415)
(113,1425)
(114,1288)
(115,1261)
(116,1160)
(117,1094)
(118,1010)
(119,942)
(120,815)
(121,785)
(122,700)
(123,686)
(124,597)
(125,521)
(126,546)
(127,451)
(128,434)
(129,343)
(130,344)
(131,289)
(132,267)
(133,226)
(134,208)
(135,204)
(136,202)
(137,157)
(138,125)
(139,101)
(140,105)
(141,85)
(142,83)
(143,61)
(144,61)
(145,50)
(146,41)
(147,35)
(148,28)
(149,18)
(150,23)
(151,22)
(152,20)
(153,24)
(154,13)
(155,7)
(156,5)
(157,3)
(158,3)
(159,4)
(160,6)
(161,3)
(162,10)
(163,1)
(164,3)
(165,2)
(167,1)
(169,1)
(170,1)
(172,1)
(173,1)
(176,2)
(183,1)
};
\end{axis}
\end{tikzpicture}

\caption{The distribution of a tournament metric for two formats}
\label{Fig6}

\end{figure}
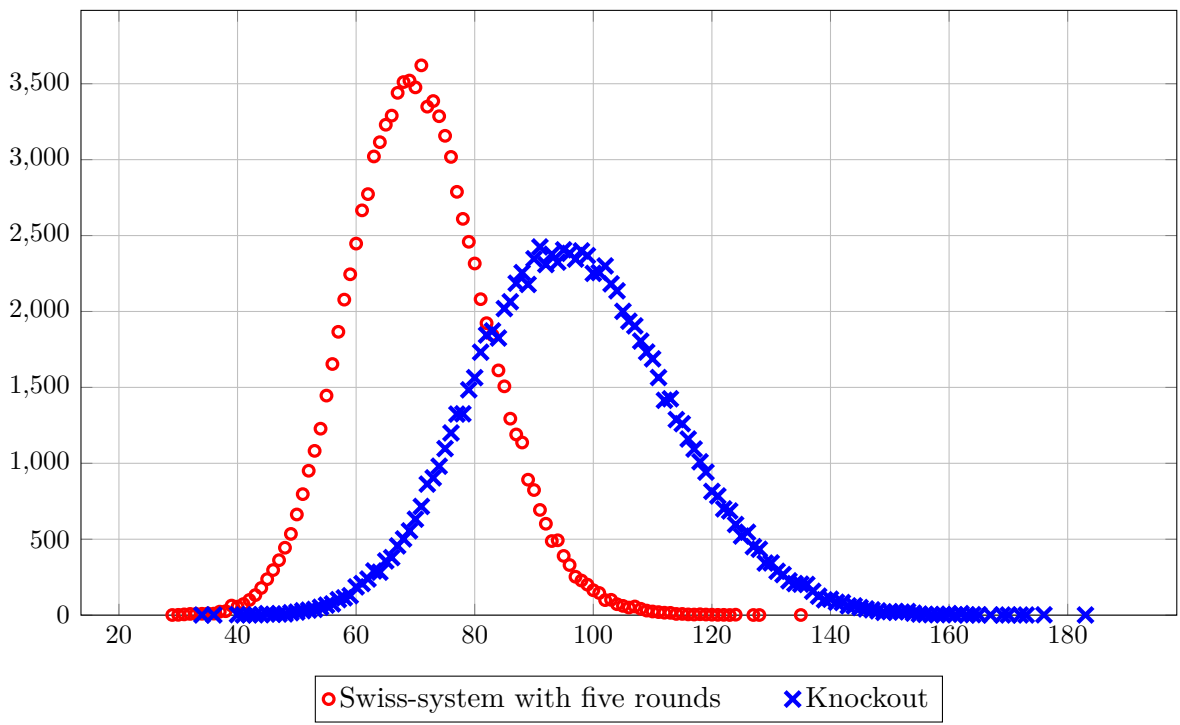


These results are based only on averages.
Figure~\ref{Fig6} presents the discrete distribution of the number of inversions for the knockout format and the Swiss-system with five rounds in the model $\texttt{skill}=5$ (based on 100 thousand simulation runs). Since the distribution can be well approximated by the normal distribution, the averages reliably describe the efficacy of the tournaments. 

\begin{table}[t!]
  \centering
  \caption{The probability that the Swiss-system with five rounds leads to \\ a ranking having a smaller number of inversions than the knockout}
  \label{Table3}
    \rowcolors{3}{}{gray!20}
    \begin{tabularx}{0.4\textwidth}{LC} \toprule 
    Model & Probability \\ \bottomrule 
    $\texttt{skill}=1$ & 0.6350 \\
    $\texttt{skill}=5$ & 0.9044 \\
    $\texttt{skill}=10$ & 0.9379 \\
    chess & 0.5598 \\
    soccer & 0.6676 \\
    tennis & 0.7095 \\ \bottomrule
    \end{tabularx}
\end{table}

Table~\ref{Table3} reports the derived estimations that the Swiss-system with five rounds outperforms the knockout design with respect to number of inversions. The advantage of the Swiss-system increases as the skill differences become more pronounced, which reinforces the message of Figures~\ref{Fig2}--\ref{Fig5}.

\input{Figure7_all_formats_inversions_standard_deviation}

The variance of the metrics can uncover how consistent the formats are in finding the true ranking. Figure~\ref{Fig7} shows the relative standard deviation of the number of inversions. The tournament designs do not differ much in this aspect but there are a couple of interesting observations. Firstly, the variance is quite high, especially when the skill differences are significant: in the $\texttt{skill}=10$ scenario, the relative standard deviation of the double round-robin format reaches almost 30\%. Secondly, increasing the number of matches slightly increases the variance. In other words, one gets more accurate but somewhat less precise results. Nevertheless, Figure~\ref{Fig7} suggests that the averages in Figures~\ref{Fig2}--\ref{Fig5} reliably reflect the efficacy of the tournament formats. 

\subsection{Seeded tournaments} \label{Sec32}

\input{Figure8_chess_seeding}

Figure~\ref{Fig8} shows that seeding can improve the efficacy of a tournament design up to 10\% (see the formats named  ``Perfect'' in Figure~\ref{Fig8}), but only when there is reliable information on the true ranking of the players. Under realistic circumstances, when the true ranking has to be approximated, the gain is reduced to at most 3\%, thus the Swiss-system remains more efficacious than any other design (see the formats named  ``Seeded'' in Figure~\ref{Fig8}). In both cases, seeding has the highest effect on the multi-stage tournament with 8 groups. On the other hand, draw and process becomes less efficacious with seeding. The likely explanation is that the second, process branch is designed according to seeding-like information, thus seeding in the first, draw branch leads to an inferior pairing in the process branch. Finding a reasonable seeding mechanism for this tournament format remains an interesting open question. 

\subsection{Comparison with previous results} \label{Sec33}

Finally, it is worth comparing our results to earlier works in the literature. \citet{Appleton1995} focuses on the percentage of times when the best player wins and finds that the Swiss-system does not perform well in this respect, for instance, draw and process seems to be better. According to Figure~\ref{Fig4}, this does not hold in general, the Swiss-system is not dominated by the draw and process even in the ability to determine the best player. Analogously, the triple knockout format strongly improves the chances of the best player compared to the knockout if there are only eight players \citep{Appleton1995} but the gain is moderated in our case (Figure~\ref{Fig4}).

Similar to us, \citet{McGarrySchutz1997} notice that the knockout is a weak tournament in its ability to rank all players but its efficacy can be enhanced by double elimination (draw and process), and especially, with an \emph{accurate} seeding. However, seeding does not help much if the initial ranking differs from the true ranking as Figure~\ref{Fig8} reveals.

We also reinforce the results of \citet{ScarfYusofBilbao2009}: tournaments with group rounds imply a higher correlation of the pre-tournament to exit ranks than knockout tournaments but this has to be traded off against the number of matches. Therefore, the conclusions of \citet{ScarfYusofBilbao2009} probably hold for a much larger set of winning probabilities.
To conclude, while our results do not contradict the main findings of previous studies, we refine important details, especially with the incorporation of the Swiss-system into the analysis. 

The lessons of our study can be summarised as follows:
(1) the Swiss-system is more efficacious than any other tournament formats containing the same number of matches, with the possible exception of the average rank of the winner (Top 1) when its performance is similar to the knockout and draw and process designs, especially if one has an accurate seeding;
(2) in the ability to reproduce the true ranking, the superiority of the Swiss-system increases with the number of matches taken into consideration;
(3) draw and process, composed from two knockout modules, outperforms multi-stage tournaments for selecting the best player but not in the number of (weighted) inversions;
(4) triple knockout shows a poor performance compared to the high number of matches played.
All of the above findings are robust with respect to the derivation of the winning probabilities and the tournament metric chosen.

\section{Conclusions} \label{Sec4}

The Swiss-system has been shown to be an efficacious and reasonable alternative to traditional tournament formats.
Its advantage is probably explained by taking directly the outcomes of previous matches into consideration. This feature is similar to the knockout format but the negative effects of an unexpected loss are mitigated.

Consequently, the Swiss-system is worth adopting in further sports. The results can be especially interesting for emerging esports, where the tournament formats are not yet solidified. Given the expanding consumer base of online sports activities, these questions will continue to be relevant for quite some time. The replacement of traditional sport formats is also not unprecedented, cf.\ the design of the most prestigious club competition in European soccer, the UEFA Champions League, where a similar format will be used from the 2024/25 season \citep{UEFA2021g}.


There remain several promising research directions.
First, it remains an open question how ties affect the efficacy of the tournament designs considered.
Second, the winning probabilities used in the paper are transitive but this assumption can be relaxed \citep{ChenJoachims2016}.
Third, each player plays the same number of matches in all our ranking mechanisms. Perhaps a similar accuracy can be achieved by removing certain clashes, especially if it is sufficient to rank the top players only.
Fourth, as the Swiss-system turns out to be a competitive tournament format, connected issues such as the optimal ranking \citep{Csato2013a, Csato2017c} and the details of the pairing algorithm \citep{BiroFleinerPalincza2017, FuhrlichCsehLenzner2021} may require further investigation. 
Finally, one may abstract from the traditional tournament formats and consider any sequence of pairwise comparisons. This direction is closely related to a research line in machine learning, where the problem is to retrieve the ranking of items from minimal number of noisy comparisons, see e.g.\ \citet{RenLiuSchroff2019}. 

\section*{Acknowledgements}
\addcontentsline{toc}{section}{Acknowledgements}
\noindent
Six anonymous reviewers provided valuable comments and suggestions on earlier drafts. \\
The research was supported by the MTA Premium Postdoctoral Research Program grant PPD2019-9/2019, 
and by the Hungarian National Research, Development and Innovation Office, grant numbers K128573, K128611, and K138945. \\
Bal\'azs R.\ Sziklai is the grantee of the J\'anos Bolyai Research Scholarship of the Hungarian Academy of Sciences and the New National Excellence Program Bolyai+ scholarship of the Ministry for Innovation and Technology. \\
P\'eter Bir\'o acknowledges the financial support by the Hungarian Academy of Sciences, Momentum Grant No.~LP2021-2.

\bibliographystyle{apalike}
\bibliography{All_references}

\clearpage
\section*{Appendix}
\addcontentsline{toc}{section}{Appendix}

\renewcommand\thetable{A.\arabic{table}}
\setcounter{table}{0}

\begin{table}[ht!]
\caption{Heatmap of the winning probabilities \\
The cell $(i,j)$ shows the winning probability of player $i$ against player $j$.}
\label{Table_A1}

\begin{subtable}{0.5\textwidth}
\caption{$\texttt{skill}=1$}
\label{Table_A1a}
\centering

\scalebox{.15}{
\setlength\extrarowheight{20pt}
\begin{tabular}{lllllllllllllllllllllllllllllllll}
   & 1                            & 2                            & 3                            & 4                            & 5                            & 6                            & 7                            & 8                            & 9                            & 10                           & 11                           & 12                           & 13                           & 14                           & 15                           & 16                           & 17                           & 18                           & 19                           & 20                           & 21                           & 22                           & 23                           & 24                           & 25                           & 26                           & 27                           & 28                           & 29                           & 30                           & 31                           & 32                           \\
\end{tabular}
}
\end{subtable}
\begin{subtable}{0.5\textwidth}
\caption{Chess}
\label{Table_A1b}
\centering

\scalebox{.15}{
\setlength\extrarowheight{20pt}
\begin{tabular}{lllllllllllllllllllllllllllllllll}
   & 1                            & 2                            & 3                            & 4                            & 5                            & 6                            & 7                            & 8                            & 9                            & 10                           & 11                           & 12                           & 13                           & 14                           & 15                           & 16                           & 17                           & 18                           & 19                           & 20                           & 21                           & 22                           & 23                           & 24                           & 25                           & 26                           & 27                           & 28                           & 29                           & 30                           & 31                           & 32                           \\
\end{tabular}
}
\end{subtable}

\vspace{0.5cm}
\begin{subtable}{0.5\textwidth}
\caption{$\texttt{skill}=5$}
\label{Table_A1c}
\centering

\scalebox{.15}{
\setlength\extrarowheight{20pt}
\begin{tabular}{lllllllllllllllllllllllllllllllll}
   & 1                            & 2                            & 3                            & 4                            & 5                            & 6                            & 7                            & 8                            & 9                            & 10                           & 11                           & 12                           & 13                           & 14                           & 15                           & 16                           & 17                           & 18                           & 19                           & 20                           & 21                           & 22                           & 23                           & 24                           & 25                           & 26                           & 27                           & 28                           & 29                           & 30                           & 31                           & 32                           \\
\end{tabular}
}
\end{subtable}
\begin{subtable}{0.5\textwidth}
\caption{Soccer}
\label{Table_A1d}
\centering

\scalebox{.15}{
\setlength\extrarowheight{20pt}
\begin{tabular}{p{0.1cm}llllllllllllllllllllllllllllllll}
   & 1                            & 2                            & 3                            & 4                            & 5                            & 6                            & 7                            & 8                            & 9                            & 10                           & 11                           & 12                           & 13                           & 14                           & 15                           & 16                           & 17                           & 18                           & 19                           & 20                           & 21                           & 22                           & 23                           & 24                           & 25                           & 26                           & 27                           & 28                           & 29                           & 30                           & 31                           & 32                           \\
\end{tabular}
}
\end{subtable}

\vspace{0.5cm}
\begin{subtable}{0.5\textwidth}
\caption{$\texttt{skill}=10$}
\label{Table_A1e}
\centering

\scalebox{.15}{
\setlength\extrarowheight{20pt}
\begin{tabular}{lllllllllllllllllllllllllllllllll}
   & 1                            & 2                            & 3                            & 4                            & 5                            & 6                            & 7                            & 8                            & 9                            & 10                           & 11                           & 12                           & 13                           & 14                           & 15                           & 16                           & 17                           & 18                           & 19                           & 20                           & 21                           & 22                           & 23                           & 24                           & 25                           & 26                           & 27                           & 28                           & 29                           & 30                           & 31                           & 32                           \\
\end{tabular}
}
\end{subtable}
\begin{subtable}{0.5\textwidth}
\caption{Tennis}
\label{Table_A1f}
\centering

\scalebox{.15}{
\setlength\extrarowheight{20pt}
\begin{tabular}{lllllllllllllllllllllllllllllllll}
   & 1                            & 2                            & 3                            & 4                            & 5                            & 6                            & 7                            & 8                            & 9                            & 10                           & 11                           & 12                           & 13                           & 14                           & 15                           & 16                           & 17                           & 18                           & 19                           & 20                           & 21                           & 22                           & 23                           & 24                           & 25                           & 26                           & 27                           & 28                           & 29                           & 30                           & 31                           & 32                           \\
\end{tabular}
}
\end{subtable}
\end{table}


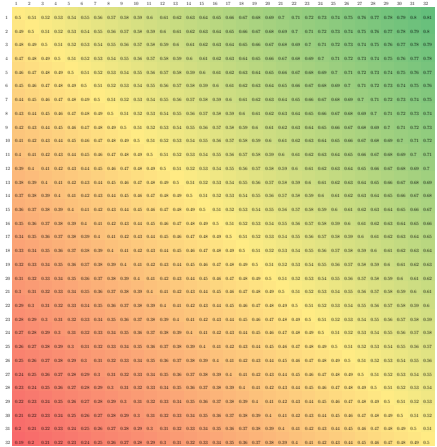
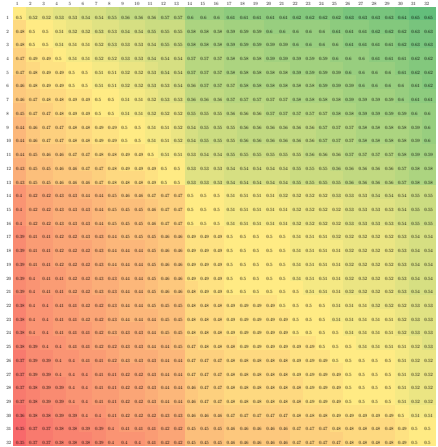
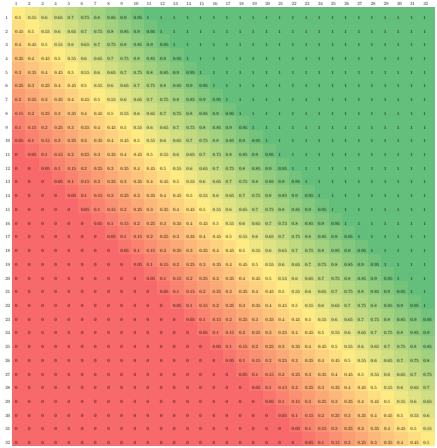
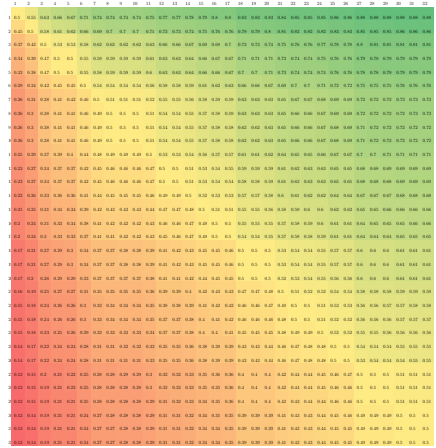
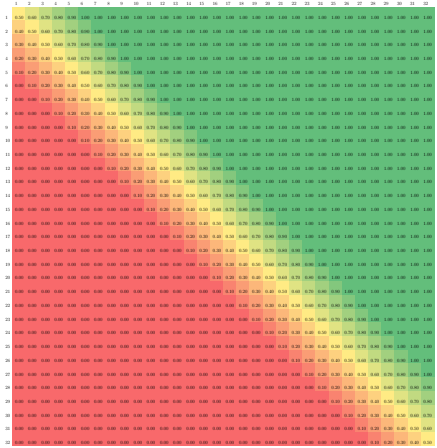
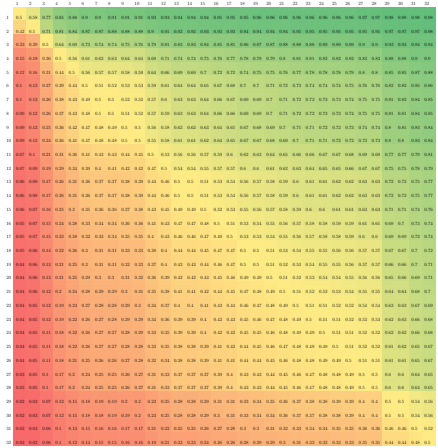
\end{document}